\newcommand \kms {{\rm km~s}^{-1}}

\newcommand \beqn {\begin{equation}}
\newcommand \eeqn {\end{equation}}
\newcommand \Planck {{\em Planck }}
\newcommand \nczhecssz {7,721 }
\newcommand \nczhecsszfast {168 }
\newcommand \nclusterhecssz {24 }
\newcommand \nclusterhecsszsdss {30 }
\newcommand \nclusterhecsplanckcomplete {83 }
\newcommand \nclusterhecsplanck {123 }

\documentclass[apj]{emulateapj}
\usepackage{epsfig}
\usepackage{natbib}
\newcommand{\noprint}[1]{}

\begin{document}

\title{HeCS-SZ: The Hectospec Survey of Sunyaev-Zeldovich Selected Clusters} 
\shorttitle{HeCS-SZ: Hectospec Survey of SZ-Selected Clusters}
\shortauthors{Rines, Geller, \& Diaferio}

\author{Kenneth J. Rines\altaffilmark{1}, Margaret J. Geller\altaffilmark{2}, 
Antonaldo Diaferio\altaffilmark{3,4}, and Ho Seong Hwang\altaffilmark{5}} 
\email{kenneth.rines@wwu.edu}

\altaffiltext{1}{Department of Physics \& Astronomy, Western Washington University, Bellingham, WA 98225; kenneth.rines@wwu.edu}
\altaffiltext{2}{Smithsonian Astrophysical Observatory, 60 Garden St, Cambridge, MA 02138}
\altaffiltext{3}{Universita' di Torino, Dipartimento di Fisica, Torino, Italy}
\altaffiltext{4}{Istituto Nazionale di Fisica Nucleare (INFN), Sezione di Torino, Torino, Italy}
\altaffiltext{5}{School of Physics, Korea 
Institute for Advanced Study, 85 Hoegiro, Dongdaemun-Gu, 130-722 Seoul, 
Korea}

\begin{abstract}

We estimate cluster masses 
and velocity dispersions for \nclusterhecsplanck  clusters from optical 
spectroscopy to compare the Sunyaev-Zeldovich (SZ) 
mass proxy and dynamical masses. 
Our new survey, HeCS-SZ (Hectospec Cluster Survey of SZ-selected clusters), 
includes \nczhecssz new or remeasured redshifts from MMT/Hectospec observations
of \nclusterhecssz 
SZ-selected clusters at redshifts $z$=0.05-0.20 and not in previous surveys.
We supplement the Hectospec data with spectra from the Sloan Digital Sky Survey
(SDSS) and cluster data from the Cluster Infall Regions in SDSS (CIRS) project and 
the Hectospec Cluster Survey (HeCS), our Hectospec survey of clusters selected by X-ray flux.
We measure the scaling relation between velocity dispersion and SZ mass estimates
from the integrated Compton parameter for an SZ complete sample of 
\nclusterhecsplanckcomplete clusters.  The observed relation agrees very well 
with a simple virial scaling from mass (based on SZ)
to velocity dispersion.  The SZ mass estimates 
(calibrated with hydrostatic X-ray mass estimates) are not significantly biased. 
Further, the velocity dispersion of cluster galaxies is consistent with the 
expected velocity dispersion of dark matter particles, indicating that 
galaxies are good dynamical tracers (i.e., velocity bias is small).
Significant mass bias in SZ mass estimates could relieve tension between cosmological 
results from \Planck SZ cluster counts and \Planck CMB data.
However, the excellent agreement between our measured velocity dispersions
and those predicted from a virial scaling relation suggests that 
any SZ mass bias is too small to reconcile SZ and CMB results.
In principle, SZ mass bias and velocity bias of galaxies 
could conspire to yield good agreement, but the required velocity
bias is $\sigma_{galaxy}\approx 0.77\sigma_{DM}$, outside the range
of plausible models of velocity bias in the literature.

\end{abstract}


\section{Introduction}

As the universe evolves, the comoving number density
of clusters of fixed mass increases.  The evolution of cluster abundances 
depends strongly on the amount of dark matter and dark energy in the 
universe. Thus, many groups have used different cluster mass proxies to 
determine the mass function and constrain cosmological parameters \citep[e.g.,][and references
therein]{cirsmf,rines08,vikhlinin09b,henry09,mantz10a,rozo10}.  
Recently, others have used the Sunyaev-Zeldovich (SZ)
effect \citep[][]{sz72} to identify large samples of clusters 
to constrain cosmological parameters \citep{benson13,hasselfield13,planckszcatalog,planckszcosmo}.

Data from the {\em Planck} satellite
show that cosmological parameters determined from anisotropies in 
the cosmic microwave background disagree with those derived from 
cluster abundance measurements from the {\em Planck} SZ cluster 
survey \citep{planckszcosmo,planckszcosmo2015}. Fewer clusters are observed than
predicted by the cosmology that best fits the \Planck CMB data.  
Interestingly, estimates of the amplitude of structure from cosmic shear yield 
a similar tension with \Planck CMB data \citep{maccrann15}.  If SZ masses 
(calibrated from X-ray observations) systematically
underestimate true masses by about 45\%,
the cosmological parameters derived from SZ cluster counts 
shift into agreement with the CMB results  \citep{planckszcosmo}. 
An alternate analysis using weak lensing data for mass calibration 
finds no significant tension \citep{vonderlinden14b,mantz15},
suggesting that the tension could arise from biases in the 
calibration of SZ masses.

Here, we compare SZ mass estimates to dynamical mass estimates 
based on the redshifts of cluster members.  Dynamical mass estimates 
have a long history beginning with \citet{zwicky1933,zwicky1937}.  
In numerical simulations, either the virial theorem or the caustic technique 
can provide cluster mass estimates with little bias but with some intrinsic 
scatter due to projection effects \citep{diaferio1999,evrard07,serra11,mamon13,gifford13,old14}.
Hydrodynamical simulations show that the velocity distribution of galaxies 
is very similar to that of dark matter particles \citep{faltenbacher06,lau10}, with the possible
exception of the brightest few galaxies \citep{lau10,wu13}.  Thus, virial masses, 
caustic masses, or dynamical mass proxies such as velocity dispersion 
are a powerful test of SZ mass estimates.

\citet{hecsysz} made the first comparison of SZ signals to mass 
estimates from galaxy dynamics, but the sample was limited to 15 clusters.
A later study by \citet{sifon12}
obtained optical spectroscopy for 16 SZ-selected clusters selected from observations
with the Atacama Cosmology Telescope (ACT); they found that the scaling relation 
between SZ signal and mass (actually measured from velocity dispersions)
is consistent with relations determined with other mass calibrators (X-ray, lensing). 
\citet{ruel14} measured velocity dispersions for SZ-selected clusters identified 
in observations with the South Pole Telescope (SPT); they conclude that SZ signal 
correlates well with velocity dispersion.
The SPT results \citep{bocquet15} are consistent with positive velocity
bias (that is, the velocity dispersion of the galaxies is larger than the velocity 
dispersion of the dark matter particles).
The clusters in the ACT and SPT samples span a wide range of redshifts ($0.2<z<1.3$). 
It is possible that the scaling between velocity dispersion and virial mass
evolves significantly over that period.  Further, the spectroscopy for these 
clusters is often incomplete at large radii or contains relatively few cluster members. 
In principle, the measured velocity dispersions could be biased \citep[e.g.,][]{biviano06,wu13}. 

To provide a much broader foundation for comparison of dynamical and SZ 
mass proxies, we compare SZ mass estimates of \nclusterhecsplanck clusters from the {\em Planck} SZ catalog 
with velocity dispersions from wide-field optical spectroscopy.   Several clusters 
have redshifts in the Sloan Digital Sky Survey \citep[][]{sdssdr10}, and many are 
part of the Cluster Infall Regions in SDSS project \citep[CIRS;][]{cirsi} or the 
Hectospec Cluster Survey \citep[HeCS;][]{hecsultimate}.  To supplement this sample and 
create an SZ-selected sample of clusters, we conducted HeCS-SZ, an MMT/Hectospec 
spectroscopic survey of \nclusterhecssz clusters.  We also include analysis of 
\nclusterhecsszsdss clusters from SDSS redshifts. 

We discuss the cluster samples and spectroscopic data in $\S 2$.  We measure
the SZ-optical scaling relations in $\S 3$. We discuss
the implications of our results in the context of other cosmological observations
in $\S 4$.  
We assume a cosmology of
$\Omega_m$=0.3, $\Omega_\Lambda$=0.7, and $H_0$=70 km s$^{-1}$ Mpc$^{-1}$ for all calculations.

\section{Observations}

\subsection{Optical Photometry and Spectroscopy}

HeCS-SZ is an extension of the HeCS survey to include 
clusters that enable construction of an SZ-limited sample. We observed 
\nczhecssz new redshifts in \nclusterhecssz clusters. We combine these 
new measurements with the existing HeCS and CIRS surveys 
and with data from the literature to construct a total sample of  
\nclusterhecsplanck clusters. 
For all but a few clusters the sampling is sufficient for a robust 
determination of velocity dispersion.
We use SDSS photometry for all clusters. 

\subsubsection{Spectroscopy: CIRS and HeCS}

The Hectospec Cluster Survey (HeCS) is a spectroscopic survey of 
58 galaxy clusters at moderate redshift ($z$=0.1-0.3)
with  MMT/Hectospec.  
HeCS includes all
clusters with ROSAT X-ray  fluxes of $f_X>5\times10^{-12}$erg s$^{-1}$ at [0.5-2.0] keV from the Bright Cluster Survey \citep[BCS;][]{bcs} or REFLEX survey \citep{reflex} with optical imaging 
in the Sixth Data Release (DR6) of SDSS \citep{dr6}.
We used DR6 photometry to select Hectospec targets.
The HeCS targets are all brighter than $r$=20.8 (SDSS catalogs are 95\%
complete for point sources to $r$$\approx$22.2).

For HeCS, we acquired spectra with the
Hectospec instrument \citep{hectospec} on the MMT 6.5m telescope.
Hectospec provides simultaneous spectroscopy of up to 300 objects
across a diameter of 1$^\circ$.  This telescope and instrument
combination is ideal for studying the virial regions and outskirts of
clusters at these redshifts.  Because cluster properties such as 
projected velocity dispersion depend on radius, wide-field spectroscopic coverage is 
important for measuring accurate global velocity dispersions 
and virial masses \citep{biviano06}.
We used the red sequence to preselect likely
cluster members as primary targets, and we filled otherwise unassigned fibers with bluer
targets \citep[][describes the details of target selection]{hecsultimate}.

CIRS used spectroscopy from the Fourth Data Release of SDSS to 
study the virial and infall regions of clusters.  We use the dynamical 
data tabulated in CIRS for 25 clusters.  We update dynamical parameters 
for two additional CIRS clusters: A2249 was poorly sampled in DR4 but has many 
more redshifts available in DR10.  We use the DR10 redshifts to update
the dynamical parameters.  The central region of A2175 was poorly sampled 
in DR4. We thus obtained additional redshifts in the central parts of A2175 
with Hectospec (see below). 

\begin{figure} 
\plotone{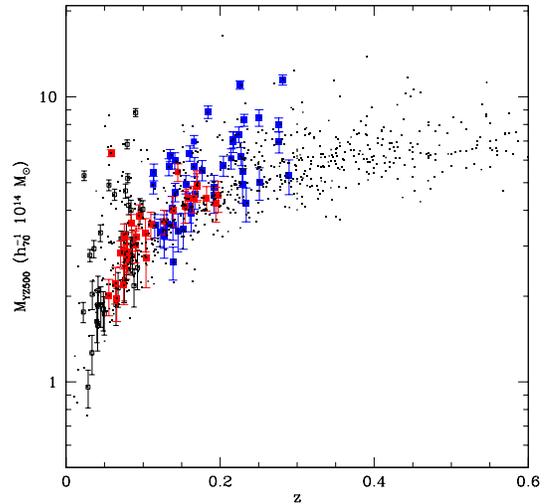}
\caption{\label{hecsplanckmz} {\em Planck} SZ mass estimates versus redshift.  
Open black squares show clusters with dynamical mass estimates from CIRS.
Solid blue and red squares show clusters from HeCS and new HeCS-SZ clusters respectively. 
Small points show the remainder of the {\em Planck} SZ catalog.  The clusters studied
here are representative of clusters at $z$$<$0.3 in the {\em Planck} SZ catalog. }
\end{figure}

\begin{figure} 
\plotone{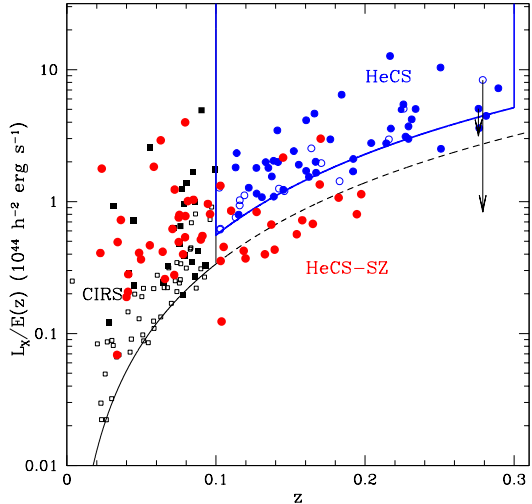}
\caption{\label{hecsplancklxz} {\em ROSAT} X-ray luminosities of {\em Planck}-selected 
clusters versus redshift.  Filled symbols are clusters in the HeCS-SZ sample: 
black squares show clusters with dynamical mass estimates from CIRS, blue 
points are clusters from HeCS, and red points are new clusters in HeCS-SZ.
}
\end{figure}

\subsubsection{Spectroscopy: HeCS-SZ}

We observed \nclusterhecssz clusters in the {\em Planck} catalog of SZ clusters using 
MMT/Hectospec.  The target clusters are in the redshift 
range $0.05\le z \le 0.20$ and were observed mostly in decreasing 
{\em Planck} signal-to-noise ratio (a few clusters with relatively 
weak SZ signals were observed as backup targets for variable observing 
conditions). 
We also observed one field in A2175, a cluster from 
CIRS with limited SDSS spectroscopy in CIRS.  Preliminary analysis 
indicated that A2175 had an unusually small velocity dispersion given its 
SZ mass.  The additional redshifts in A2175 show that the CIRS data led to a 
significant underestimate of its velocity dispersion and caustic mass. 

Our observing strategy closely matches that of HeCS: we used SDSS 
photometry to identify a red sequence in each cluster field.  We then identify
a cutoff in apparent magnitude that offers a good compromise of high 
completeness (sparser targets produce fewer fiber conflicts) and dense 
sampling.  Targets are primarily drawn from galaxies with $g-r$ colors within 
0.2 mag of the red sequence, and we assign higher priorities to brighter
galaxies and galaxies closer to the cluster center.  This approach provides 
reasonably high sampling in the cluster cores but can lead to relatively sparse 
sampling of dense regions outside the core. We included galaxies with slightly 
bluer colors (up to 0.4 mag bluer than the red sequence) as targets to fill fibers 
when available.  We matched all targets to redshifts from the literature as compiled 
by NED\footnote{http://ned.ipac.caltech.edu} as of 2013 September as well as to SDSS DR8 spectra. Most 
of the targets with existing redshifts are from SDSS, but several are from targeted 
studies of individual clusters \citep[e.g.,][for A586]{cypriano05}.
Targets with existing redshifts are removed from the targeting catalogs prior 
to fiber assignment.  

Table \ref{hecsredshifts} lists \nczhecssz new redshifts measured 
with Hectospec.  We visually inspected all spectra to confirm the reliability of 
the redshift.  Column 5 of Table \ref{hecsredshifts} lists the cross-correlation
score $R_{XC}$ from the IRAF package {\em rvsao} \citep{km98}.  A score of 
$R_{XC}>3$ indicates a reliable redshift; some galaxies with smaller values 
of $R_{XC}$ are included when visual inspection shows multiple obvious 
absorption and/or emission lines and the spectrum suffers from contamination
(e.g., light bleeding into the spectrum from a nearby fiber containing a bright star). 
Table \ref{hecsmemredshifts} lists redshifts from SDSS 
and other literature (as compiled by NED) for galaxies classified as cluster 
members by the caustic technique (see below).  
Table \ref{hecsszfast}
lists \nczhecsszfast redshifts measured with the FAST instrument \citep{fast} on the 
1.5-meter Tillinghast telescope at the Fred Lawrence Whipple Observatory.  
The additional single-slit spectra from FAST reduce the incompleteness 
of bright (SDSS $r\lesssim 16.5$) galaxies in the HeCS-SZ clusters.

In addition, we identified several clusters in the {\em Planck} SZ catalog that lie below the 
completeness limits but that are at sufficiently low redshift ($z\lesssim 0.1$) that they 
have reasonable redshift coverage in SDSS DR10.  We include these 
clusters in an extended sample. 

We include four nearby ($z\le 0.05$) clusters that lie inside the SDSS DR10 photometric footprint but 
outside the SDSS spectroscopic footprint.  These nearby clusters 
have large numbers of redshifts available in the literature.  Because of the redshift 
dependence of the limiting mass for SZ detection by \Planck~(driven by the large beam 
size of {\em Planck}), including these low-redshift clusters improves the sampling of 
low-mass clusters in the sample.  The FAST redshifts in Table \ref{hecsszfast} are 
especially useful for these clusters. 

Figure \ref{hecsplanckmz} shows the \Planck SZ mass estimates versus redshift. 
The minimum mass a cluster must have to 
be detected by \Planck increases with redshift because the SZ signal of lower-mass 
clusters at higher redshift is diluted by the large beam below the sensitivity of {\em Planck}. 

The CIRS and HeCS clusters provide a good sampling of the $M_{SZ}-z$ distribution,
but this distribution is possibly biased due to the underlying X-ray selection of 
CIRS and HeCS. 
Figure \ref{hecsplancklxz} shows the X-ray luminosity of clusters in CIRS, HeCS, 
and HeCS-SZ as a function of redshift.  The clusters we target with Hectospec include clusters 
that lie above the X-ray flux limits of CIRS and HeCS but were not in the 
appropriate SDSS photometric footprint and also clusters that have X-ray fluxes
below the CIRS/HeCS flux limits. 
Targeting these X-ray-faint clusters enables a test of the impact of X-ray selection 
on the scaling relation parameters based on SZ and optical properties.
The X-ray luminosities are measured in the ROSAT band but from heterogeneous
sources \citep{bcs,noras,bohringer05,piffaretti11,planckszcatalog}.  A careful
study of the X-ray properties of HeCS-SZ clusters would require a homogeneous 
reanalysis of ROSAT X-ray images.

\begin{figure} 
\plotone{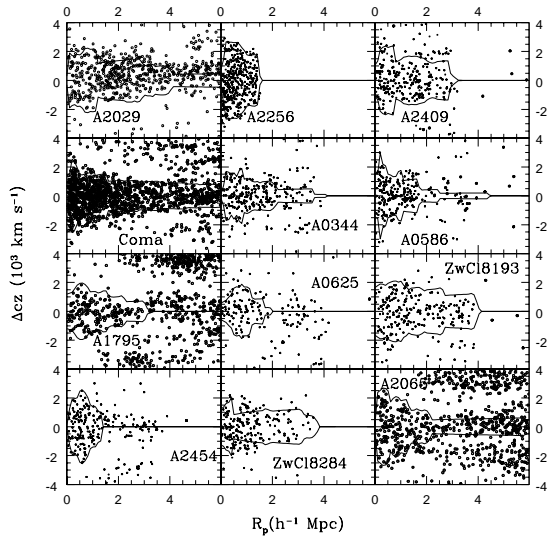}
\caption{\label{hecsplanckc1} {Redshift (rest-frame clustrocentric velocity) versus projected radius for galaxies around
HeCS-SZ clusters.  The caustic
pattern is evident as the trumpet-shaped regions with high density.
The solid lines indicate our estimate of the location of the caustics
in each cluster.  Clusters are ordered left-to-right and top-to-bottom
by decreasing mass as estimated from the \Planck ~SZ data. }}
\end{figure}

\begin{figure} 
\plotone{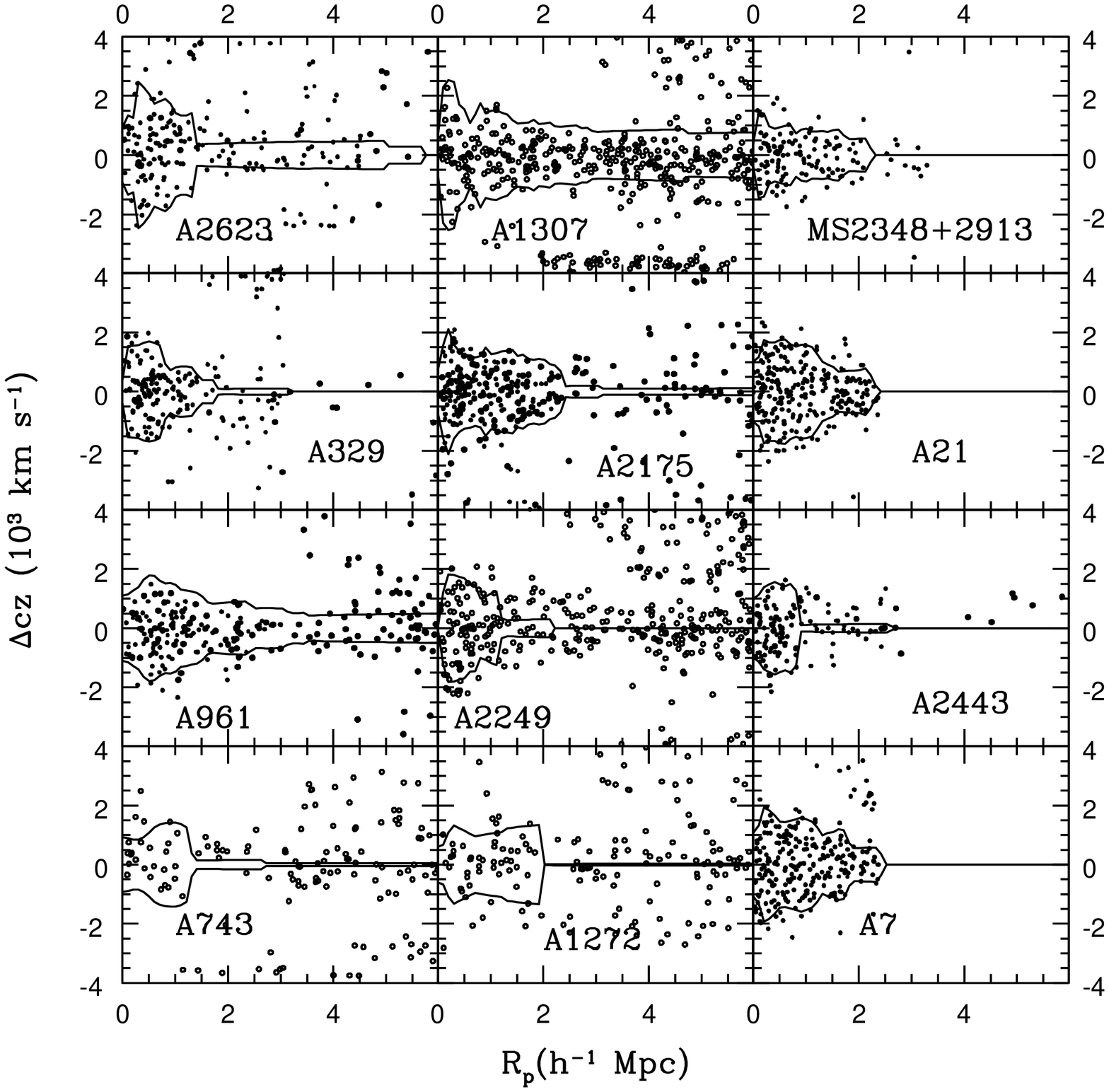}
\caption{\label{hecsplanckc2} {Same as Figure \ref{hecsplanckc1}. }}
\end{figure}

\begin{figure} 
\plotone{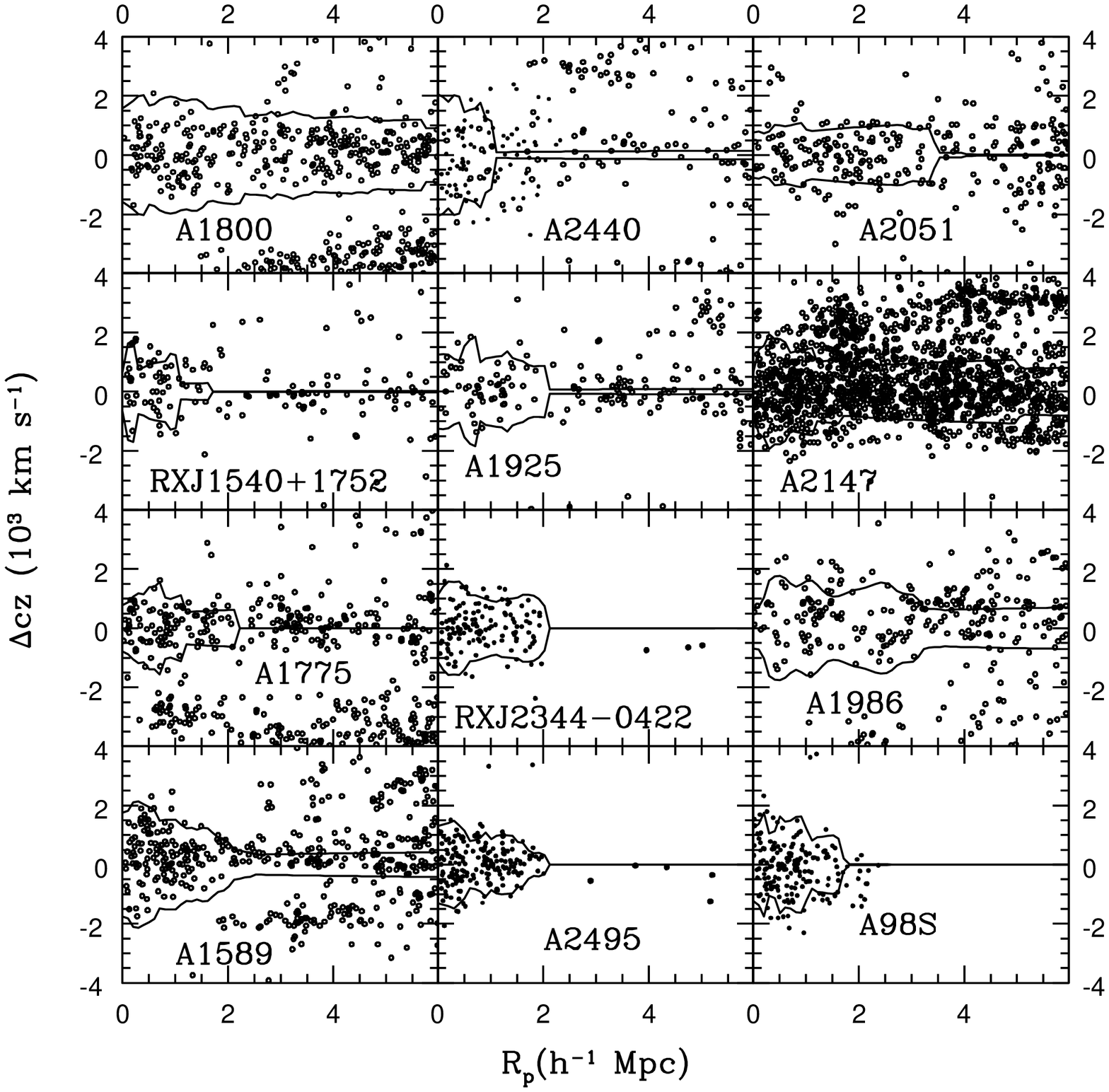}
\caption{\label{hecsplanckc3} {Same as Figure \ref{hecsplanckc1}.   }}
\end{figure}

\begin{figure} 
\plotone{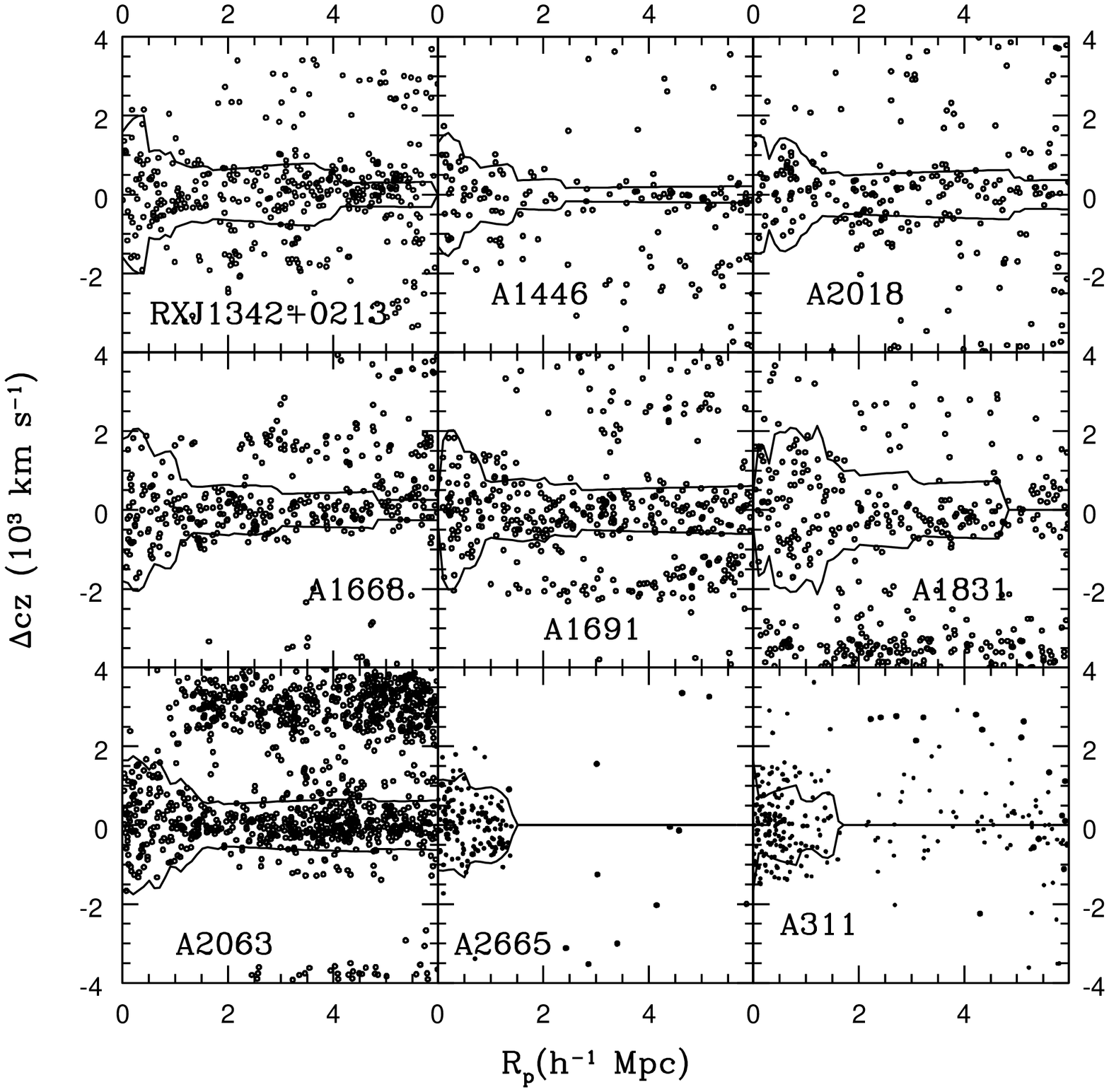}
\caption{\label{hecsplanckc4} {Same as Figure \ref{hecsplanckc1}.  }}
\end{figure}

\begin{figure} 
\plotone{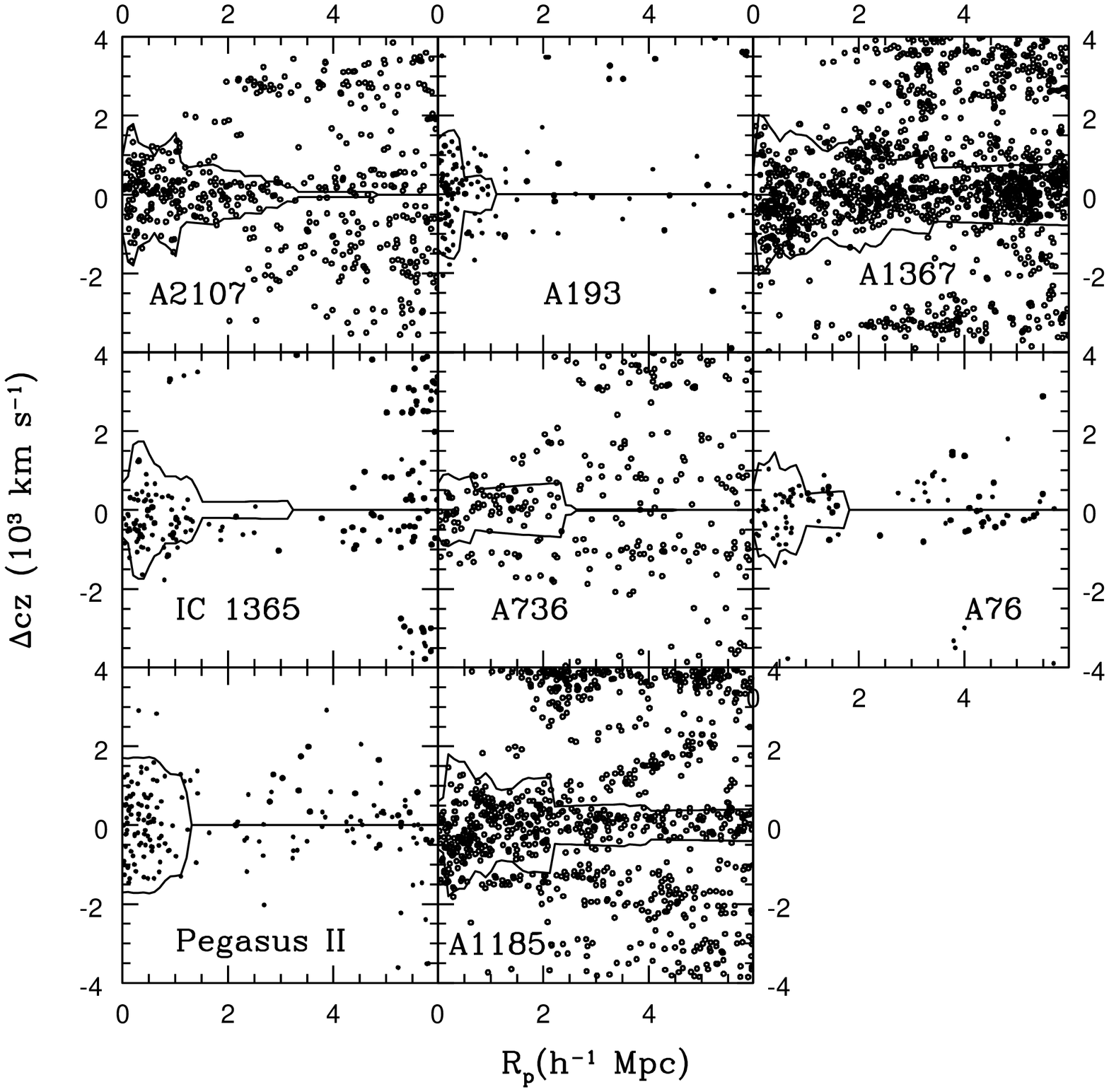}
\caption{\label{hecsplanckc5} {Same as Figure \ref{hecsplanckc1}. }}
\end{figure}

\begin{table}[th] \footnotesize
\begin{center}
\caption{\label{hecsredshifts} \sc HeCS-SZ Redshifts from MMT/Hectospec}
\begin{tabular}{ccrrccc}
\tableline
\tableline
\tablewidth{0pt}
\multicolumn{2}{c}{Coordinates (J2000)} & $cz_\odot$ & $\sigma_{cz}$  & $R_{XC}$ & Flag & Member \\ 
 RA  & DEC  &  km/s &km/s & &  & \\ 
\tableline
00:09:33.60    & 32:31:03.16   &  83460   &     52.19  & 6.23  & Q       &0\\
00:09:35.55    & 32:14:05.00   &  69478   &    182.17  & 1.64 &  Q      & 0\\
00:09:39.32    & 32:21:22.38   & 122251  &      49.89 &  5.08 &  Q    &   0\\
00:09:42.73    & 32:16:05.45   &  83482         & 100      & 4.45  & Q       &0\\
00:09:43.80    & 32:33:54.17   & 108270  &       9.69  &15.29 &  Q   &    0\\
\tableline
\end{tabular}
\tablecomments{Table \ref{hecsredshifts} is published in its entirety in the electronic edition of the Journal. A portion is shown here for guidance regarding its form and content.}
\end{center}
\end{table}

\begin{table}[th] \footnotesize
\begin{center}
\caption{\label{hecsmemredshifts} \sc HeCS-SZ Members from Literature Redshifts }
\begin{tabular}{ccrrc}
\tableline
\tableline
\tablewidth{0pt}
\multicolumn{2}{c}{Coordinates (J2000)} & $cz_\odot$ & $\sigma_{cz}$  & Ref. \\
 RA  & DEC  &  km/s &km/s & \\ 
\tableline
0:11:45.24   &  32:24:56.17    & 30309 & 100 &2 \\
0:11:19.72    & 32:17:09.39    & 32168 & 201 & 2 \\
0:20:02.98    & 28:44:58.73   &  29876 & 27 & 2 \\
0:20:05.48    & 28:41:01.73   &  29545 & 47 & 2 \\
0:20:16.85    & 28:46:09.69   &  26793 & 33 & 2 \\
\tableline
\end{tabular}
\tablecomments{Table \ref{hecsmemredshifts} is published in its entirety in the electronic edition of the Journal. A portion is shown here for guidance regarding its form and content.}
\tablecomments{References: [1] SDSS, [2] NED. }
\end{center}
\end{table}

\begin{table}[th] \footnotesize
\begin{center}
\caption{\label{hecsszfast} \sc HeCS-SZ Redshifts from FLWO 1.5m/FAST}
\begin{tabular}{ccrrcc}
\tableline
\tableline
\tablewidth{0pt}
\multicolumn{2}{c}{Coordinates (J2000)} & $cz_\odot$ & $\sigma_{cz}$  & $R_{XC}$ & Member \\ 
 RA  & DEC  &  km/s &km/s & \\ 
\tableline
0:11:05.08  &    31:54:29.53 &          24574   &     24 &  9.01 & 0\\
0:11:34.79  &    32:28:16.28 &         30990    &    20  & 17.46 & 1 \\
0:11:45.24  &    32:24:56.20 &         30542    &    51  & 6.47  &1 \\
0:12:27.58  &    32:45:09.84 &          12600   &     9   & 15.40 & 0 \\
0:12:30.47  &    32:19:12.45 &          24565   &     5   & 31.67 & 0 \\
\tableline
\end{tabular}
\tablecomments{Table \ref{hecsszfast} is published in its entirety in the electronic edition of the Journal. A portion is shown here for guidance regarding its form and content.}
\end{center}
\end{table}

\begin{figure*} 
\epsscale{1.0}
\plotone{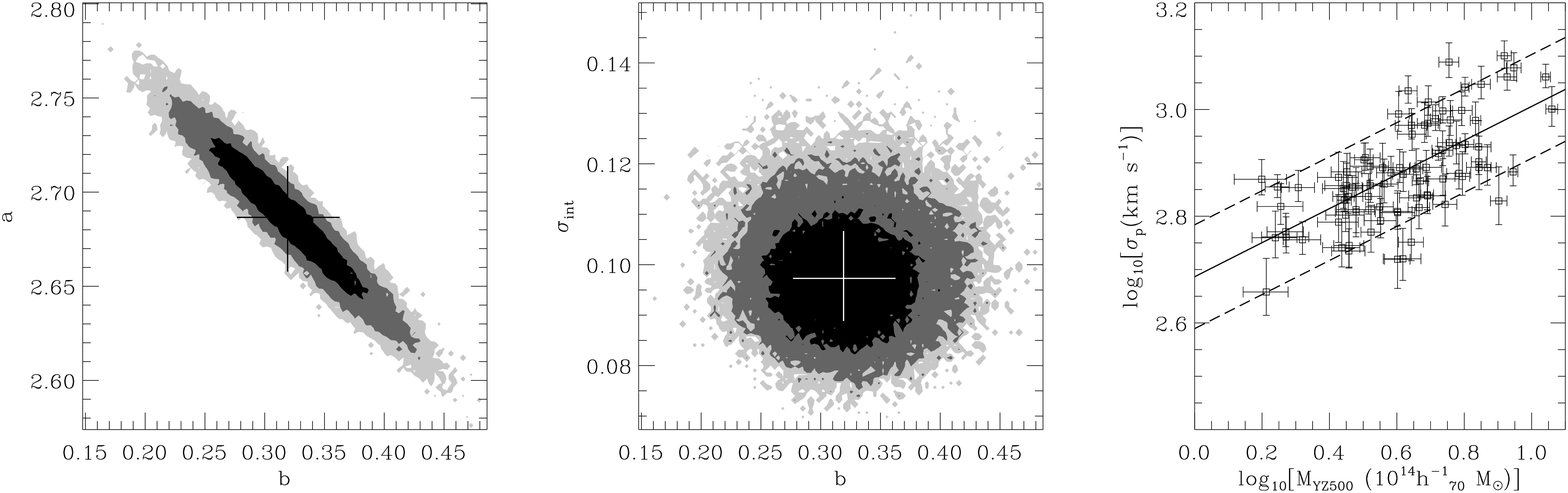}
\caption{\label{sigmapvsmysz} Scaling relation between projected velocity 
dispersion $\sigma_p$ and $M_{SZ}$,
the mass proxy based on the integrated Compton parameter $Y_{SZ}$.
The first two panels
show the marginalized 
probability distribution functions of the parameters $a$ (intercept), 
$b$ (slope), and $\sigma_{int}$ (the intrinsic scatter).
The first panel shows that the slope and intercept are correlated. 
Shaded regions indicate 68\%, 95\%, and 99\% confidence regions 
from darkest to lightest shading. 
The crosses show the median and the 68\% confidence range of
the probability distribution functions. The third panel shows 
the the relation with the
median values of $a$ and $b$ as a solid line with the median value of the intrinsic scatter $\sigma_{int}$ shown by the dashed lines. }
\end{figure*}

The caustic technique \citep{diaferio1999,serra11} uses a redshift-projected radius 
diagram to isolate cluster members  from foreground and background galaxies 
in phase space.  After smoothing the galaxy distribution in the redshift diagram, the infall regions of 
clusters produce well-defined envelopes containing the vast majority of cluster 
members.  Specifically, the list of cluster members within $r_{200}$ is 96\% complete and only 2\% of the members are actually interlopers; within the larger
radius $3r_{200}$ , where the caustic technique is the only usable method, the completeness is 95\% and the interloper fraction is 8\% \citep{serra12}.
The edges of this distribution are called caustics and they are related 
to the escape velocity profile of the cluster \citep[see][for  reviews]{diaferio09,serra11}.  
The escape velocity profile is the basis for a mass profile that can extend 
into the infall region where the galaxies are gravitationally bound but not virialized.
Caustic mass estimates generally agree with
estimates from X-ray observations and gravitational lensing
\citep[e.g.,][and references therein]{cairnsi,bg03,diaferio05,cirsi,cirsmf,hecslens}.

Figures \ref{hecsplanckc1}-\ref{hecsplanckc5} show the phase space 
diagrams of the HeCS-SZ clusters not already published in CIRS or 
HeCS (the poorly-sampled CIRS clusters A2175 and A2249 are reproduced 
here with enlarged datasets).  Almost all clusters display prominent infall 
patterns, and the caustics are shown on the figures.
Clusters are ordered by decreasing SZ mass,
and there is a clear trend of decreasing central velocity dispersion 
with decreasing SZ mass.

We apply the prescription of \citet{danese} to determine the mean
redshift $cz_\odot$ and projected velocity dispersion $\sigma_p$ of
each cluster from all galaxies within the caustics.  We calculate
$\sigma_p$ using only the cluster members projected within $r_{200}$
estimated from the caustic mass profile.  
Note that our measured velocity dispersions use the caustic technique 
only to define membership and the limiting radius $r_{200}$.  Independent
of its performance as a mass estimator, the caustic technique is a highly 
efficient membership selection algorithm, especially at the relatively 
small radii we focus on here \citep{serra12}.  Table \ref{hecsysztab} 
lists the central cluster redshifts, velocity dispersions inside $r_{200}$,
and $M_{200}$ from the caustic mass profile. The ninth column of 
Table \ref{hecsysztab} indicates whether the cluster is part of the CIRS, 
HeCS, or HeCS-SZ sample.

\subsection{SZ Measurements}

The SZ measurements are from \citet[][]{planckszcatalog}, an all-sky SZ 
survey.
Numerical simulations indicate that the integrated Compton y-parameter
$Y_{SZ}$ has smaller scatter than the peak y-decrement $y_{peak}$
\citep{motl05,planckszcatalog}. \citet{planckszcatalog} report only $Y_{SZ}$. Although
$y_{peak}$ should be nearly independent of redshift, $Y_{SZ}$ depends
on the angular size of the cluster.  The quantity $Y_{SZ}D_A^2$
removes this dependence.   Table \ref{hecsysztab} summarizes the 
{\em Planck} SZ measurements.

The {\em Planck} mass estimates are extracted from an aperture of 
$\theta_{500}$, the angular radius corresponding to $r_{500}$ (the radius 
$r_\Delta$ is the radius that encloses a mean density of $\Delta \rho_c(z)$ 
where $\rho_c(z)$ is the critical density).  This radius is larger than the radii 
probed by some other mass estimators.  For instance, \citet{bonamente08}  
and \citet{mantz10b} find 
that X-ray masses are best determined within $r_{2500}$ 
\citep[although ][and others use $M_{500}$]{vikhlinin09a}.  
\citet{marrone09} uses an aperture of 350 kpc as the best match to their 
mass estimates from strong gravitational lensing.  Because the SZ signal 
falls off more slowly with radius than the X-ray flux, the outer parts of 
clusters are more important for SZ observables than for X-ray observables.  
For instance, \citet{plancklargeradii} used {\em Planck} data to determine 
the average pressure profile of the ICM to radii of 3$r_{500}$, a regime 
that is very difficult to study even with very deep {\em Chandra} 
observations.  Because virial masses and velocity dispersions are best 
suited for mass estimates at radii  $\sim r_{200}$, they may be better 
suited for comparison with SZ mass estimates. 

The central redshifts in the \Planck SZ catalog are usually close to the central 
redshifts we obtain in our hierarchical clustering analysis of the cluster 
redshifts (see D99 for details).  However, for about half of the clusters,
our central redshifts differs by more than a percent from the redshifts listed in the \Planck 
SZ catalog.  We therefore re-scale all SZ integrated Compton parameters 
by $[D_A^2(z_h)/D_A^2(z_{SZ})]$
where $D_A(z_h)$ and $D_A(z_{SZ})$ are the angular diameter distances 
for the hierarchical center $z_h$ and the \Planck catalog redshift $z_{SZ}$.
We similarly rescale SZ mass estimates using the appropriate scaling relation
from \citet{planckszcatalog}.

We define an SZ-complete sample of \nclusterhecsplanckcomplete  clusters from the 
SZ mass proxy $M_{500}$ in the \Planck SZ catalog.  The final column of Table 
\ref{hecsysztab} indicates whether the cluster is in the complete sample.  The completeness 
limit corresponds to the 80\% completeness limit for the medium-deep survey
covering 44\% of the sky and to the 50\% completeness limit for the shallow 
survey covering the remaining 56\% of the sky \citep{planckszcatalog}.
Our sample includes all but four clusters above this limit: two at moderate
redshift (A1677 at $z$=0.18 and A1759 at $z$=0.17) and two at low redshift
($z\approx 0.04$: A2572 and RBS 1929). 
The SZ completeness limits we use are slightly above the 80\% completeness 
limits of the updated \Planck SZ catalog \citep{planckszcatalogv2}.  A quick inspection shows 
that the updated SZ catalog contains few clusters above the completeness
limits we use here.

\begin{table*}[th] \footnotesize
\begin{center}
\caption{\label{hecsysztab} \sc  Dynamical Masses and SZ Signals}
\begin{tabular}{lcccccccccc}
\tableline
\tableline
\tablewidth{0pt}
Cluster & $\alpha$ & $\delta$ & $z$    & $\sigma_p$ & $M_{200,c}$ & $M_{SZ}$ & $Y_{SZ}D_A^2$  & Spectra & \Planck ID & Sample \\ 
 & $\deg$ & $\deg$ &    & $\kms$ & $10^{14} M_\odot$ & $10^{14} M_\odot$ & $10^{-5}$Mpc$^{-2}$ \\ 
\tableline
A0007    & 2.93500 &  32.41700 &0.10302 & $783^{+58}_{-48}$   & 2.77   $\pm$ 1.14 &  $3.317^{+0.420}_{-0.456}$ &   $0.105^{+0.025}_{-0.024}$  & HeCS-SZ & PSZ1G113.26-29.69 & 1 \\
A0021    & 5.17050 &  28.67510 &0.09456 & $761^{+54}_{- 44}$   & 2.92  $\pm$ 1.33 &   $3.825^{+0.359}_{-0.376}$ &   $0.146^{+0.025}_{-0.025}$  & HeCS-SZ & PSZ1G114.78-33.72  & 1 \\
A0076   & 10.00200  &  6.81800 &0.03999 & $455^{+66}_{- 46}$   & 1.19 $\pm$  0.04  &   $1.631^{+0.243}_{-0.258}$ &   $0.032^{+0.009}_{-0.008}$ & HeCS-SZ & PSZ1G118.03-55.88  & 1\\
A0085   & 10.45870  & -9.30190 &0.05565 & $692^{+55}_{- 45}$   & 2.50 $\pm$  1.19 &   $4.900^{+0.213}_{-0.217}$ &   $0.225^{+0.018}_{-0.018}$ & CIRS & PSZ1G115.20-72.07  & 1\\
A0098S &    11.61470 &  20.38645 &0.10380 & $594^{+48}_{- 39}$ &  2.17 $\pm$  0.09&  $2.733^{+0.516}_{-0.591}$ &   $0.079^{+0.029}_{-0.028}$ & HeCS-SZ & PSZ1G121.35-42.47 & 0 \\
\tableline
\end{tabular}
\end{center}
\tablecomments{Table \ref{hecsysztab} is published in its entirety in the electronic edition of the Journal. A portion is shown here for guidance regarding its form and content.}
\tablecomments{Redshift $z$ and velocity dispersion $\sigma_p$ are computed for galaxies defined as members using the caustics.   }
\end{table*}

\section{Results}

As discussed in $\S 1$, data from the \Planck satellite indicate 
tension between cosmological parameters determined from CMB 
and SZ results.  One possible resolution to the tension is that the SZ mass
estimates (calibrated with hydrostatic X-ray mass estimates) are biased.
Comparing dynamical estimates of
cluster mass from galaxy redshift surveys to the SZ mass proxies tests this hypothesis. 
Several studies show a strong correlation between X-ray mass estimates
and SZ mass estimates \citep[e.g.,][]{bonamente08,andersson11,planckearlycalibration,czakon15}, but both methods measure the 
properties of the intracluster medium (ICM). Thus, systematic effects could still be 
present.  For instance, the ICM is likely to depart from hydrostatic equilibrium 
in the outer parts of the cluster \citep{bonamente13}.  
Gravitational lensing does not measure the ICM, but it does measure all of the
matter along the line of sight to the cluster, introducing significant scatter 
into lensing mass estimates \citep[e.g.,][]{hoekstra01,hoekstra11b,hwang14}.  \citet{marrone09} 
show that lensing masses are consistent with SZ estimates although with 
significant scatter. 
Recently, \citet{vonderlinden14b} and \citet{hoekstra15} have used large 
samples of weak lensing mass estimates to test for systematic bias in 
SZ masses; both groups find that the SZ masses are systematically 
underestimated, but both estimates of bias are smaller than the value 
required to fully reconcile \Planck CMB and SZ results \citep[the uncertainty 
range in bias obtained by][includes this value within the 2$\sigma$ confidence interval]{vonderlinden14b}. 
In contrast, \citet{melin15} use weak lensing of the CMB to estimate cluster masses, 
and they find little evidence for mass bias.

Figure \ref{sigmapvsmysz} shows the relation between projected velocity dispersion 
$\sigma_p$ and the mass $M_{SZ}$ estimated from the {\em Planck} 
data (note that, for most clusters, the measurements of $Y_{SZ}$ use 
X-ray data to determine the region where the SZ signal is extracted).  
We use a Bayesian approach (see Appendix for details) to determine the 
best-fit relation $P(\sigma_p|M_{SZ})$, that is, the predicted value of 
$\sigma_p$ at a given observed value of $M_{SZ}$.  We allow for 
intrinsic scatter in $\sigma_p$ that is expected to arise from projection 
effects of non-spherical clusters. Our Bayesian analysis yields a relation of 
\beqn
log_{10}(\sigma_p) = 0.319^{+0.043}_{-0.042}log_{10}(M_{YSZ}) + 2.687^{+0.027}_{-0.029}
\eeqn
with $\sigma_p$ in units of $\kms$ and $M_{SZ}$ in units of $10^{14} h_{70}^{-1} M_\odot$.  
The scatter in $\sigma_p$ at fixed $M_{SZ}$ is $\log_{10}\sigma=0.0973^{+0.0094}_{-0.0085}$.  
The best-fit parameters and their uncertainties are the medians and the boundaries of the 
68\% confidence levels derived from the posterior probability of the regression parameters.
Figure \ref{sigmapvsmysz} shows this relation as a solid line. 
Note that we fit $P(\sigma_p|M_{SZ})$ rather than the inverse because the 
statistical uncertainties in $M_{SZ}$ are smaller than the statistical uncertainties in $\sigma_p$.

The intrinsic scatter we measure corresponds to about a factor of two in 
the estimated mass within $r_{200}$.  A comparison of several richness-based 
and dynamics-based mass estimators demonstrate similar scatter for
several mass estimators based on velocity dispersions or variations of Jeans' 
analysis \citep{old13}.  Thus, the scatter probably represents geometric 
projection effects and not our use of the caustic technique to define
cluster membership.

\begin{figure} 
\epsscale{1.0}
\plotone{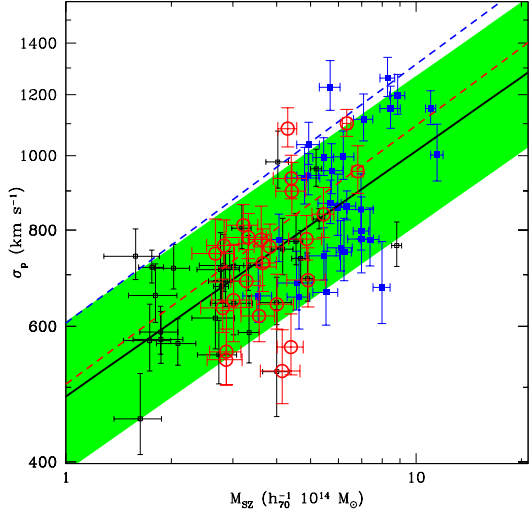}
\caption{\label{sigmapvsmysz3} Scaling relation between projected velocity dispersion $\sigma_p$ and 
the SZ mass proxy $M_{SZ}$ based on the integrated Compton parameter $Y_{500}D_A^2$. 
The thick solid line shows the best-fit relation of 
$P(\sigma_p|M_{SZ})$ with the intrinsic scatter shown as the green band.  
Open squares, filled squares, and open circles represent clusters from CIRS, 
HeCS, and HeCS-SZ respectively.  The dotted line shows the 
relation from \citet{hecsysz} from a small number of clusters.  The red dashed line shows the predicted 
relation using the virial scaling relation from \citet{evrard07} and assuming no hydrostatic
mass bias (i.e., 1-$b$=0).  The blue dashed line shows the same predicted relation for 
the bias (1-$b$=0.58) required to reconcile \Planck ~SZ counts with the \Planck ~CMB cosmology.
}
\end{figure}

\begin{figure} 
\epsscale{1.0}
\plotone{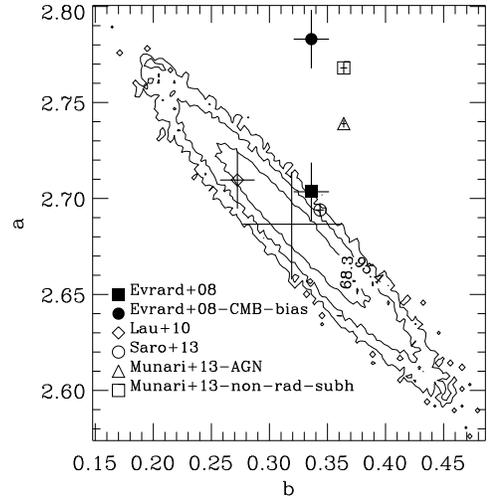}
\caption{\label{biasmodels} Parameters of the virial scaling relation between projected velocity dispersion $\sigma_p$ and 
the SZ mass proxy $M_{SZ}$ based on the integrated Compton parameter $Y_{500}D_A^2$ ($a$ is the intercept, $b$ is the slope). 
Contours show confidence intervals from our Bayesian analysis and the cross without a symbol shows the median
and 68\% percentiles of the distribution shown with the contour levels.  Points with errorbars show models based on simulations. 
The filled square is the virial scaling relation of dark matter particles from \citet{evrard07}.  The filled circle shows this same 
relation re-normalized to reflect a mass bias of $M_{SZ}=0.59M_{true}$, the value needed to match SZ and CMB constraints. 
The other points show 
several models of velocity bias.   The open triangle and open square show the models
of \citet{munari13} for galaxies identified from dark matter subhalos and from hydrodynamical 
simulations including star formation and AGN feedback.  The open diamond shows the model of \citet{lau10}, and the open circle 
shows the model of \citet{saro13}. 
}
\end{figure}

Previous work provides an expected value for this slope.  Numerical 
simulations of clusters with a variety of codes yield a consistent scaling relation 
of the mass $M_{200}$ with velocity dispersion, $\sigma_p\propto M_{200}^{0.33}$ \citep{evrard07}.
This slope is measured for randomly selected dark matter particles rather than 
galaxies, but hydrodynamical simulations suggest that velocity bias is small 
for large samples of cluster galaxies \citep{wu13} like
HeCS and CIRS (we discuss velocity bias further in $\S \ref{velocitybias}$).
The slope of the scaling relation for dark matter particles in clusters agrees
well with our observed $\sigma_p-M_{SZ}$ relation (Equation 1). 
Figure \ref{sigmapvsmysz3} shows our data and scaling relation
compared to the virial scaling of dark matter particles, and the 
agreement is reasonable.  Figure \ref{biasmodels} shows the marginalized 
probability distribution functions of the parameters of our scaling
relation along with the virial scaling of dark matter particles.
Figures \ref{sigmapvsmysz3} and \ref{biasmodels} also show the virial scaling 
of dark matter particles rescaled by assuming that $M_{SZ}=0.58 M_{true}$, 
the mass bias required to match the SZ counts to the CMB data.  
Such a large mass bias is strongly disfavored by our observations.

\begin{figure*} 
\epsscale{1.0}
\plotone{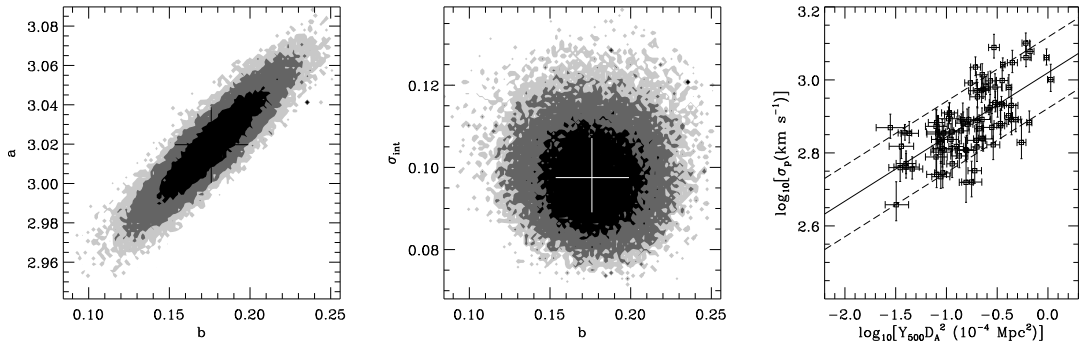}
\caption{\label{sigmapvsysz} Similar to Figure \ref{sigmapvsmysz} for the scaling relation between projected velocity 
dispersion $\sigma_p$ and  the integrated Compton parameter $Y_{SZ}$.}
\end{figure*}

Figure \ref{sigmapvsysz} shows the best-fit relation for $P(\sigma_p|Y_{SZ}D_A^2)$, 
the expected velocity dispersion at fixed SZ mass proxy $Y_{SZ}D_A^2$.  
\citet{planckszcatalog} obtain $M_{500}^{1.79} \propto (Y_{SZ}D_A^2)$ using 
hydrostatic mass estimates from detailed XMM-Newton observations.  Because the 
concentration-mass relation depends weakly on mass \citep[e.g.,][]{bullock01}, we 
use a fixed conversion of $M_{200}\approx 1.35M_{500}$ appropriate for 
concentration $c$=5 assuming an NFW profile \citep{nfw97}.  With these assumptions,
the expected slope of the $\sigma_p-Y_{SZ}$ relation is 0.188 with an intercept of 3.003.  
Figure \ref{sigmapvsysz} shows that these values agree very well with our
Bayesian analysis (see also Table \ref{hecsyszfits}).

In contrast with our previous work \citep{hecsysz}, the relation between projected 
velocity dispersion $\sigma_p$ and $Y_{SZ}D_A^2$ agrees with expectations 
from scaling relations of dark matter particles and simulations of the SZ effect. 
We attribute this difference to both the much larger (5x) sample of clusters 
studied here and the improved statistical methods enabled by the 
larger sample.

\begin{figure*} 
\epsscale{1.0}
\plotone{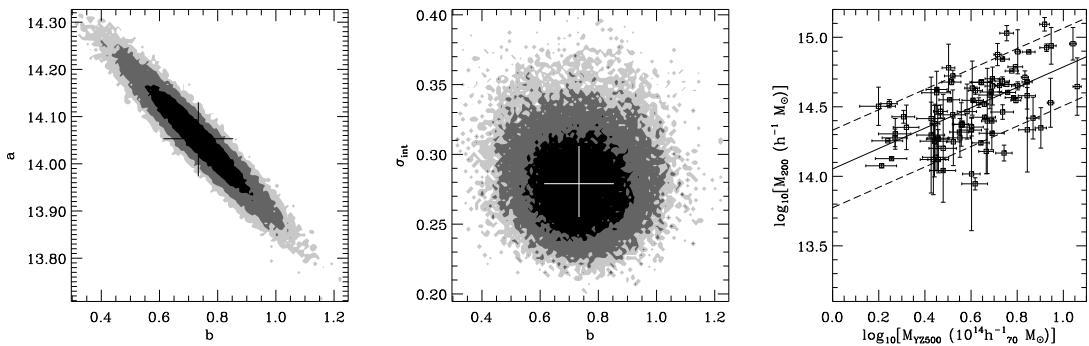}
\caption{\label{m200vsmysz} Similar to Figure \ref{sigmapvsmysz} for the scaling relation between caustic mass $M_{200}$  
and $M_{SZ}$,
the mass proxy based on the integrated Compton parameter $Y_{SZ}$.}
\end{figure*}

Figure \ref{m200vsmysz} shows the best-fit relation $P(M_{200}|M_{SZ})$, the 
caustic mass $M_{200}$ obtained at fixed $M_{SZ}$.  The intrinsic  scatter in 
this relation is somewhat smaller than a factor of two, consistent with the 
expected scatter in caustic mass estimates due to projection effects \citep{serra11}.
Note that a similar level of scatter is found for alternate implementations of the 
caustic technique \citep{gifford13} as well as alternative mass estimators based 
on measured velocity dispersions \citep{old14}.
While a detailed treatment of outliers is beyond the scope of this work, we note that 
one cluster, MS2348+2929, with an observed velocity dispersion smaller than 
predicted by its \Planck SZ mass, is undetected in observations with the Arcminute 
Microkelvin Imager \citep[][]{amiplanck2015}, suggesting that the SZ mass
in the \Planck catalog is an overestimate.

\begin{figure} 
\epsscale{1.0}
\plotone{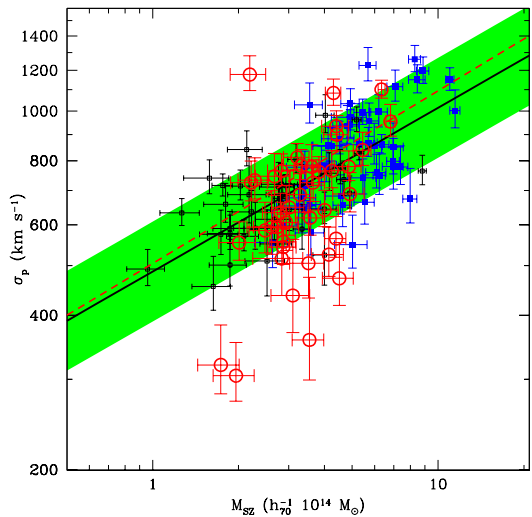}
\caption{\label{sigmapvsmyszext} Scaling relation between projected velocity dispersion $\sigma_p$ and 
the SZ mass proxy $M_{SZ}$ based on the integrated Compton parameter $Y_{500}D_A^2$
for the extended sample of \Planck-selected clusters (including clusters below the 
\Planck ~completeness limits). Several clusters in the extended sample are 
outliers below the main relation.  These clusters are not well sampled in 
SDSS spectroscopy, so their velocity dispersions are likely underestimated.
}
\end{figure}

Figure \ref{sigmapvsmyszext} shows the $\sigma_p-M_{SZ}$ relation for the extended sample 
of \nclusterhecsplanck clusters.  There are significantly more outliers than in the 
SZ-complete sample.  Most of these outliers have redshifts only from SDSS, and some 
are at $z>0.1$.  Thus, these clusters are not well sampled.  Obtaining additional redshifts for these
clusters could significantly alter the measured velocity dispersions (similar to the
changes for the CIRS clusters A2175 and A2249 resulting from additional redshift data from Hectospec and SDSS respectively).  The best-fit parameters of the 
scaling relation are virtually unchanged, but the inferred intrinsic scatter is larger
due to the larger number of outliers (Table \ref{hecsyszfits}).

\begin{table*}[th] \footnotesize
\begin{center}
\caption{\label{hecsyszfits} \sc  Scaling Relations Between Dynamical Masses and SZ Signals}
\begin{tabular}{lccc}
\tableline
\tableline
\tablewidth{0pt}
Relation & $b$   & $a$ & $\sigma_y$ \\
\tableline
$P(\sigma_p|Y_{500}D_A^2)$ & $0.176^{+0.023}_{-0.022}$ & $3.020^{+0.019}_{-0.019}$ & 0.0975$^{+0.0096}_{-0.0084}$ \\
extended sample & $0.191^{+0.022}_{-0.022}$ & $3.023^{+0.021}_{-0.021}$ & 0.1182$^{+0.0096}_{-0.0088}$ \\
CIRS/HeCS & $0.175^{+0.029}_{-0.030}$ & $3.013^{+0.024}_{-0.024}$ & 0.114$^{+0.012}_{-0.010}$ \\
$P(Y_{500}D_A^2|\sigma_p)$ & $2.36^{+0.31}_{-0.29}$ & $-7.57^{+0.83}_{-0.91}$ & $0.371^{+0.035}_{-0.031}$ \\
extended sample & $2.02^{+0.29}_{-0.26}$ & $-6.63^{+0.75}_{-0.83}$ & $0.394^{+0.029}_{-0.026}$ \\
\tableline
$P(M_{200}|M_{SZ})$ & $0.73^{+0.12}_{-0.12}$ & $14.053^{+0.077}_{-0.080}$ & $0.279^{+0.027}_{-0.024}$ \\
extended sample & $0.76^{+0.12}_{-0.12}$ & $14.006^{+0.072}_{-0.071}$ & $0.346^{+0.027}_{-0.024}$ \\
CIRS/HeCS & $0.70^{+0.15}_{-0.15}$ & $14.069^{+0.099}_{-0.096}$ & $0.308^{+0.034}_{-0.029}$ \\
$P(M_{SZ}|M_{200})$ & $1.72^{+0.57}_{-0.79}$ & $-24.2^{+11.5}_{-8.3}$ & $0.46^{+0.26}_{-0.19}$ \\
\tableline
$P(\sigma_p|M_{SZ})$ & $0.319^{+0.043}_{-0.041}$ & $2.687^{+0.027}_{-0.029}$ & $0.0973^{+0.0094}_{-0.0085}$ \\
extended sample & $0.339^{+0.043}_{-0.041}$ & $2.665^{+0.025}_{-0.026}$ & $0.1198^{+0.0096}_{-0.0087}$ \\
$P(M_{SZ}|\sigma_p)$ & $1.42^{+0.16}_{-0.19}$ & $-3.47^{+0.55}_{-0.47}$ & $0.205^{+0.019}_{-0.017}$ \\
extended sample & $1.03^{+0.12}_{-0.09}$ & $-2.36^{+0.26}_{-0.34}$ & $0.222^{+0.017}_{-0.015}$ \\
\tableline
\end{tabular}
\end{center}
\tablecomments{Fits are of the relation $P(y|x)$ assuming 
the linear form $\log{y} = a + b\log{x}$ with intrinsic scatter $\sigma_{\log{y}}$ in the relation at fixed values of $\log{x}$.}
\end{table*}

\section{Discussion}

\subsection{Predictor Relations}

Cluster scaling relations applied to
large surveys are a basis for cosmological studies, including measuring the 
cluster mass function or the power spectrum \citep[][]{mantz10a,rozo10}.  \citet{andreon10} 
discusses how, given observable properties A and B, the slopes of the 
predictor relation $P(A|B)$ (the probability of a cluster having the property A 
given an observed value of property B) may be significantly different from the 
inverse of the slope of the predictor relation $P(B|A)$.  This difference is larger
when there is significant intrinsic scatter in the relation between the two properties.

Because different investigators require different predictor relations, we 
include here the relations between several mass observables (Table \ref{hecsyszfits}). 
We do not include constraints on the $M_{SZ}-M_{200}$ relation for the 
extended sample because the large scatter caused by a few outliers leads 
to very weak constraints on the parameters of the scaling relation. 

\subsection{Impact on the Tension Between {\em Planck} Cosmological Parameters from SZ versus CMB}

Cosmological constraints from {\em Planck} observations of the CMB 
predict a higher normalization of the cluster mass function 
(parameterized by $\Omega_m$ and $\sigma_8$) than the measured abundance from
the SZ cluster detections  \citep{planckszcosmo}.  As discussed 
in that paper, the cluster constraints are based on a scaling relation 
between SZ integrated Compton decrement and X-ray masses (calculated 
with the assumption of hydrostatic equilibrium).  They assume a 
hydrostatic mass bias due to non-thermal pressure support parameterized 
as $M_{true}=(1-b)M_{HSE}$ where $M_{true}$ and $M_{HSE}$ are respectively the 
true cluster mass and the mass estimated under the assumption of hydrostatic equilibrium.
Hydrodynamic simulations of intracluster gas \citep[][]{nagai07,nelson14} predict a value of $(1-b)=0.8$, 
and the cosmological constraints are derived by allowing this parameter to vary 
in the range $0<b<0.3$.   
The tension between the SZ and CMB constraints can be eliminated by assuming that the hydrostatic
mass bias is significantly larger, $(1-b)=0.58\pm0.04$ \citep{planckszcosmo,planckszcosmo2015}.   
Note that the parameter $b$ can have non-zero values either because of
non-thermal pressure support or because of other calibration 
offsets \citep[e.g., XMM-Newton temperature calibration, see][]{israel15,schellenberger15}.

Estimates of hydrostatic mass bias from comparisons of X-ray and lensing mass 
estimates find smaller offsets \citep[e.g.,][]{vikhlinin09a,mahdavi13,applegate14}.
Recent revisions to systematic uncertainties in lensing mass estimates yield 
consistency in mass estimates of individual clusters between different 
investigators \citep{vonderlinden14b,hoekstra15}, yielding estimates of 
$(1-b)\approx 0.7-0.8$, intermediate between no hydrostatic mass bias 
and the large bias required to match CMB constraints. Alternatively, a new
method of measuring weak lensing of the CMB by clusters yields
$1/(1-b)=0.99\pm0.19$, consistent with little to no mass bias \citep{melin15}.

It is thus very interesting to see whether our dynamical mass estimates imply 
small hydrostatic mass bias (leaving tension between clusters and the CMB)
or large hydrostatic mass bias (alleviating tension between clusters and the 
CMB but aggravating tension among different cluster mass estimators).
As mentioned in $\S 3$, our best-fit scaling relation is consistent with the 
\Planck~ scaling relation based on hydrostatic mass estimates from XMM-Newton 
observations.  Furthermore, Figure \ref{sigmapvsmysz3} shows that renormalizing 
this relation by assuming a hydrostatic mass bias of $(1-b)=0.58\pm0.04$ overpredicts 
the velocity dispersion at fixed $Y_{SZ}$ by an amount larger than the statistical 
uncertainties.  That is, the CMB cosmological parameters predict significantly 
larger cluster velocity dispersions than our measured values. 

We next consider three possible explanations of the tension between the 
CMB normalization of the cluster mass scale and our measurement of the 
relation between velocity dispersion and integrated SZ decrement.  
First, we investigate whether X-ray selection (used for part of the sample at
larger redshift) significantly impacts the resulting scaling relation.  Second, 
we discuss the possibility of velocity bias (galaxies moving faster or 
slower than dark matter particles).  Third, we discuss the possible impact
of massive neutrinos producing a smaller cluster abundance for a fixed 
matter power spectrum.

\subsection{Impact of Cluster Selection}

Scaling relations can be sensitive to the method of sample selection.  The relation 
between dynamical mass and SZ signal could depend on whether the cluster sample
is selected from an optical catalog, an X-ray catalog, or a SZ catalog, and whether the
samples are flux-limited (detection-limited) or volume-limited. For instance, clusters with 
luminous cooling cores could be overrepresented in a flux-limited X-ray catalog compared 
to a mass-limited sample. 
The HeCS and CIRS cluster samples were drawn from X-ray-selected samples. 
Thus, all clusters from these samples have moderately large X-ray fluxes. The {\em Planck}
early release clusters contained several that were not previously detected in X-rays. 
Followup {\em XMM-Newton} observations of these clusters showed that they are in 
younger dynamical states than the rest of the early release clusters \citep{planckxmmfollowup}.
Thus, it is conceivable that the HeCS and CIRS clusters are not a representative sample 
of {\em Planck} clusters.  

We use our {\em Planck}-selected sample to test whether the scaling relations depend 
on the selection technique. Specifically, we fit the scaling relations based on only 
clusters from the CIRS and HeCS samples, both of which are selected by X-ray flux. 
There is no significant change in the best-fit parameters for the X-ray selected sample
compared to the SZ-selected sample (Table \ref{hecsyszfits}\footnote{Because $M_{SZ}$ and $Y_{SZ}D_A^2$ are closely related, we do not include a separate fit for the $\sigma_p-M_{SZ}$ relation for the CIRS/HeCS subsample.}).  Thus, the impact of X-ray selection versus SZ 
selection appears to be small, at least for the large and complete samples that we consider here.

\subsection{\label{velocitybias} Can Velocity Bias Resolve the Tension?}

In numerical simulations, the velocity dispersion of randomly selected
dark matter particles closely traces the mass of dark matter halos \citep{evrard07}.
Observationally, one measures the velocity dispersion of galaxies, which 
may move faster or slower than the underlying dark matter distribution. 
This ``velocity bias" can be parametrized as $b_v=\sigma_{gxy}/\sigma_{DM}$,
where $\sigma_{gxy}$ and $\sigma_{DM}$ are the velocity dispersions 
of galaxies and dark matter particles respectively. 

If one assumes that the \Planck CMB cosmological parameters are correct, 
then the offset between the scaling relation we observe and the relation 
predicted by the CMB-based parameters provides information about the  
relation between galaxy dynamics and true cluster mass.  In particular, 
significant negative velocity bias ($b_v\approx 0.77$) is required to bring the
results into agreement.  

Modeling velocity bias in simulations is a very challenging problem, due 
to both the uncertain physics in galaxy formation and evolution and the large
dynamic range required to simulate individual cluster galaxies in a 
cosmological simulation.  Some simulations follow the evolution of 
dark matter subhalos, but the evolution of galaxies may differ significantly
because galaxies are expected to form at the centers of dark matter 
subhalos and thus survive even after their dark matter halos are tidally 
stripped.

Earlier, we used the consistency of the virial mass function of X-ray selected 
clusters with cosmological constraints from WMAP5, supernovae, and 
baryon acoustic oscillations to conclude that velocity 
bias is small: $\sigma_{gxy}=(1.05\pm0.05)\sigma_{DM}$ \citep{rines08}.
At present, there is no general agreement on the amount or even the sign
of velocity bias, but the large negative velocity bias required for consistency 
with \Planck CMB-based parameters is not predicted by any current models. 
Simulations suggest that small samples of cluster galaxies restricted only to the 
brightest members could be subject to negative velocity bias of $\sim$15\% \citep{old13,wu13},
but these simulations also suggest that large samples such as the ones we analyze 
here should not be subject to significant velocity bias.

Note that a recent analysis of the redshift-space correlation function of 
high-mass galaxies from the SDSS Baryon Oscillation Spectroscopic Survey 
(BOSS) suggests that $b_v\approx 0.86$ \citep{guo15}.  This negative 
velocity bias probably reflects the fact that even massive and rich clusters 
contain very few high-mass galaxies \citep[e.g., Figure 4 of][]{guo15}; thus, 
the measured velocity bias 
is consistent with simulations that predict negative velocity bias for 
the brightest few galaxies \citep{old13,wu13}.  It is also possible that 
the analysis of \citet{guo15} does not adequately model the impact of 
coherent infall among satellite galaxies \citep{hikage15}.  Again, the spectroscopic 
samples considered here include many galaxies below the 
characteristic absolute magnitude $M_*$ and are thus expected to 
have smaller bias than more luminous samples. 

Several recent simulations suggest that cluster galaxies should be 
positively biased by 5-15\% depending on the details of galaxy 
modeling \citep{lau10,saro13,munari13}. 
Positive velocity bias would further aggravate the tension between the velocity dispersions 
we measure and the large SZ-mass normalization required to match the CMB data.

\begin{figure} 
\epsscale{1.0}
\plotone{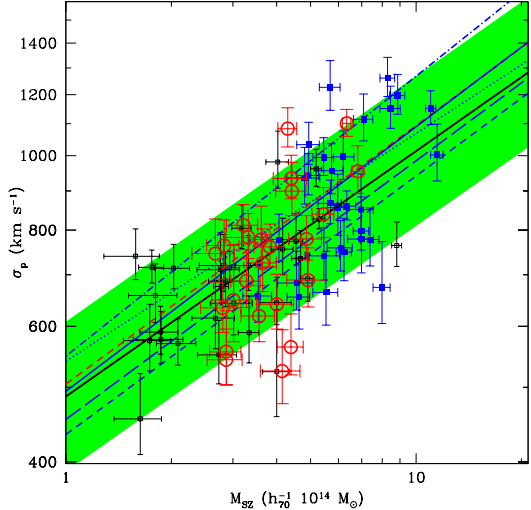}
\caption{\label{sigmapvsmyszbias} Scaling relation between projected velocity dispersion $\sigma_p$ and 
the SZ mass proxy $M_{SZ}$ based on the integrated Compton parameter $Y_{500}D_A^2$. 
The thick solid line shows the best-fit relation of 
$P(\sigma_p|M_{SZ})$ with the intrinsic scatter shown as the green band.  The other lines show 
several predictions of velocity bias.  The dash-dotted line shows the prediction 
of \citet{munari13}, the dotted line shows the prediction of \citet{lau10}, the blue 
solid line shows the prediction of \citet{saro13}, the long-dashed line shows 
the prediction from \citet{old13}, and the blue short-dashed line shows the 
velocity bias for high-mass galaxies from \citet{guo15}.  }
\end{figure}

Many of the simulations predict  that the velocity bias depends weakly on 
halo mass, so a more complete description of velocity bias may require
a virial scaling relation with arbitrary slope 
(fixed $b_v$ requires that the slope of the $\sigma_{gxy}-M$ relation
is identical to the $\sigma_{DM}-M$ relation).   Figure \ref{sigmapvsmyszbias} 
shows several of these relations compared to our data, and Figure \ref{biasmodels}
shows the parameters of some of these models compared to the uncertainties in 
our observed scaling relation.  The HeCS-SZ data and \Planck masses are 
consistent with the models of \citet{lau10}, while the models of \citet{munari13} 
lie far outside the observed relation.  Importantly, although there is no 
consensus on the exact amount of velocity bias, none of the recent estimates 
are consistent with the large velocity bias ($b_v\sim 0.77$) required to 
reconcile the \Planck SZ mass function with the CMB.  
Indeed, the 
discrepancy between our observed scaling relation and the models 
of \citet{munari13} is in the opposite direction of the discrepancy required 
to reduce the CMB-SZ tension.

\subsection{Massive Neutrinos as a Solution?}

Neutrinos with significant masses can suppress the formation of large-scale 
structure.  Thus, massive neutrinos provide one possible explanation of the 
observed deficit of SZ clusters compared to the predictions from the best-fit
$\Lambda$CDM model to the \Planck CMB data. In particular, joint fits to CMB
and SZ data from \Planck yield estimates of total neutrino 
masses $\Sigma m_\nu =(0.40\pm0.21)$eV
when allowing the hydrostatic mass bias to vary between 0 and 0.3 \citep{planckszcosmo}.
Adding baryon acoustic oscillation (BAO) measurements yields an estimate of
$\Sigma m_\nu=(0.20\pm0.09)$eV \citep{planckszcosmo}.
\citet{wyman14} point out that massive neutrinos not only alleviate tension 
between \Planck CMB results and
cluster abundance measurements, they also alleviate tension 
between \Planck CMB results and local measurements of the Hubble constant. 
Similarly, \citet{maccrann15} find that a similar tension exists 
between \Planck CMB results and cosmic shear measurements; this tension 
can be partially alleviated with the introduction of a sterile neutrino. 
However, note that an alternate analysis of the cluster mass function using X-ray luminosities 
and weak lensing mass calibration yields reasonable agreement with the \Planck 
CMB results, thus implying no need for massive neutrinos \citep{mantz15}.
The good agreement between our measured velocity dispersions and 
those predicted by the \Planck SZ masses (assuming little velocity bias)
supports the mass calibration used in the \Planck SZ analysis.  Our results 
therefore support the possibility of massive neutrinos as a solution to the
CMB-SZ tension.

Experimental measurements of neutrino oscillations place a lower limit of 
$\Sigma m_\nu>0.06$eV [95\% confidence level] \citep{capozzi14}.  Constraints
from the power spectrum of the Lyman-$\alpha$ forest from BOSS observations 
yield upper limits of $\Sigma m_\nu<$0.98 eV, or $<$0.16 eV when combined 
with \Planck CMB data \citep{palanque15}.  Thus, massive neutrinos remain a 
plausible solution to the CMB-SZ tension, but 
the required masses may produce tension with constraints from the Lyman-$\alpha$ forest.

\section{Conclusions}

The {\em Planck} satellite has produced a dramatic increase in the number of 
galaxy clusters with SZ mass estimates.  Because the catalog includes many 
nearby clusters and covers the entire sky, many clusters in the {\em Planck}
catalog have existing mass estimates from galaxy dynamics.  
Here we measure \nczhecssz new redshifts in \nclusterhecssz clusters 
to obtain a large SZ-selected sample of \nclusterhecsplanck  clusters with both 
dynamical and SZ mass estimates. 
To date, this is the largest sample of clusters used to 
compare velocity dispersions and SZ mass estimates. 
We focus on a SZ-complete sample of \nclusterhecsplanckcomplete clusters.

The measured velocity dispersions agree well with the predicted velocity 
dispersions from the cluster masses in the {\em Planck} SZ catalog and the 
virial scaling relation of dark matter particles.  The cosmological parameters
based on {\em Planck} CMB observations are not consistent with the 
mass function based on masses from the {\em Planck} SZ catalog. 
One way to resolve this tension is to allow for mass bias in the SZ masses; 
large mass bias ($M_{SZ} \approx 0.58M_{true}$) is required to 
reconcile the CMB and SZ results.  Such large mass bias is strongly 
disfavored by our results. 

In principle, velocity bias could allow galaxy velocity dispersions to agree 
with the virial scaling relation for dark matter particles based on strongly 
biased SZ masses.  However, no recent estimates of the amount of 
velocity bias are consistent with the large velocity bias ($b_v \approx 0.77$)
required for this scenario.  In fact, some models of velocity bias 
have $b_v > 1$, a possibility that would further aggravate the tension 
between a possible SZ mass bias and our measured velocity dispersions. 

Departures from a standard $\Lambda$CDM cosmological model could 
resolve the tension between CMB and SZ cosmological parameter estimates. 
For example, significant neutrino masses would decrease the amplitude 
of the power spectrum on cluster scales relative to the normalization 
from the CMB \citep{planckszcosmo}.  In this scenario, {\em Planck} cluster 
masses could have little bias, and the excellent agreement between the 
measured velocity dispersions and the virial scaling relation of dark matter 
particles would require that galaxy velocity bias is small (i.e., $b_v \approx 1$).

Future work on the equilibrium dynamics of cluster galaxies can test the 
possibility of large velocity bias: if large velocity bias is present, a Jeans 
analysis should reveal that the cluster masses are larger than inferred 
by virial scaling relations (or by the caustic technique). 
Future simulations of the evolution of galaxies within clusters could test 
whether large velocity bias is plausible.  If not, our results suggest that 
the tension between cosmological parameters derived from CMB and 
SZ data may require extensions to the standard $\Lambda$CDM cosmological model. 
Observations of SZ-selected clusters at higher redshift could measure the 
evolution of cluster scaling relations and provide further insight into the 
origin of the CMB-SZ tension.

\acknowledgements

We thank Jim Bartlett and Nabila Aghanim for advice on using the \Planck SZ catalogs.  
MJG is supported by the Smithsonian Institution.
AD acknowledges support from the grant Progetti di Ateneo/CSP TO Call2 2012 0011
``Marco Polo" of the University of Torino, the INFN grant InDark, the grant PRIN 2012 
``Fisica Astroparticellare Teorica" of the Italian Ministry of University and Research.  We 
thank Susan Tokarz for reducing the spectroscopic data and Perry 
Berlind and Mike Calkins for assisting with the observations.  We also thank 
the telescope operators at the MMT and Nelson Caldwell for scheduling 
Hectospec queue observations.

{\it Facilities:} \facility{MMT (Hectospec)}, \facility{FLWO:1.5m (FAST)}

\appendix
\section{Bayesian parameter estimation}\label{sec:bayesParam}

Define the  likelihood $p(D\vert \theta,M)$ the
probability of measuring the set of data $D$
when the model $M$ is described by the set of parameters
$\theta$; the prior $p(\theta\vert M)$ is the probability that the set $\theta$ occurs.
We are interested in estimating the probability density function (PDF) of the parameters $\theta$
given our data set $D$
\begin{equation}
p(\theta\vert D, M)={ p(D\vert\theta, M)p(\theta\vert M) \over p(D\vert M)} \; . 
\end{equation}
Given the model $M$, we need to assume the likelihood $p(D\vert \theta,M)$ and the prior $p(\theta\vert M)$,
whereas $p(D\vert M)$ is a trivial normalization factor.

In this work, we are interested in describing our data with linear
correlations between pairs $(X,Y)$ of the logarithm of the observables.
In general, a number of unknown hidden variables produces a scatter
in the linear correlation $Y=a+bX$. We model this scatter with a
single parameter, the intrinsic dispersion $\sigma_{\rm int}$.
Therefore, given a measure $X_i$ with uncertainty $\sigma_{X_i}$, the probability
of measuring $Y_i$ with uncertainty $\sigma_{Y_i}$
is $p(Y_i,\sigma_{Y_i}\vert \theta,X_i,\sigma_{X_i})$, where $\theta=\{a,b,\sigma_{\rm int}\}$.
We assume the Gaussian likelihood
\begin{equation}
p(D\vert\theta,M) = \prod_i {1\over (2\pi \sigma_i^2)^{1/2}} \exp\left[-(Y_i-a-bX_i)^2 \over 2\sigma_i^2\right]  
\end{equation}
where
\begin{equation}
\sigma_i^2=\sigma_{\rm int}^2 + \sigma_{Y_i}^2 + b^2 \sigma_{X_i}^2 \; .
\end{equation}

We assume independent flat priors for both $a$ and $b$.
For the intrinsic dispersion $\sigma_{\rm int}$, which is
positive defined, we assume
\begin{equation}
p(\sigma_{\rm int}\vert M) = {\mu^r\over \Gamma(r)}  x^{r-1} \exp(-\mu x) 
\end{equation}
where $x=1/\sigma_{\rm int}^2$, and $\Gamma(r)$ is the usual gamma function.
This PDF describes a variate with mean $ r/\mu$, and variance $ r/\mu^2$.
We set $r=\mu=10^{-5}$ which guarantees an almost flat prior.

To estimate the parameter PDF $p(\theta\vert D, M)$, we perform a Markov Chain Monte Carlo (MCMC)
sampling with the code APEMoST developed by Johannes Buchner and
Michael Gruberbauer \citep{buchner11, gruberbauer09}.
We obtain a fairly complete sampling with $2\times 10^6$ MCMC iterations.
The boundaries of the parameter space were set to
$[-100,100]$ for $a$ and $b$, and $[0.01,100]$ for $\sigma_{\rm int}$.
The initial seed of the random number generator was set
with the {\tt bash} command {\tt \verb"GSL_RANDOM_SEED=$RANDOM"}.

As the three best-fit parameters $a$, $b$, and $\sigma_{\rm int}$ of the Bayesian 
analysis, we adopt the medians derived from the posterior PDF $p(\theta\vert D,M)$. 
Likewise, we adopt the boundaries of the 68\% confidence levels around the 
medians as the uncertainties on these best-fit parameters.

\bibliographystyle{apj}
\bibliography{rines}

\end{document}